\documentclass[useAMS,usenatbib]{mn2e}

\usepackage[]{longtable}
\usepackage[]{natbib}
\usepackage{amssymb,amsmath}

\newcommand{\f}{RRAT\,J1819--1458}

\newcommand{\be}{\begin{equation}}
\newcommand{\ee}{\end{equation}}
\newcommand{\bdm}{\begin{displaymath}}
\newcommand{\edm}{\end{displaymath}}

\def\aj{\ref@jnl{AJ}}                   
             
\def\apj{{ApJ}}                 
\def\apjl{{ApJ}}

\def\aap{{A\&A}}

\def\mnras{{MNRAS}}

\def\nat{{Nature}}

\title[The extended X--ray emission around \f]{The extended X--ray emission around \f}
\author[A. Camero-Arranz, et al.]{A. Camero-Arranz$^{1}$\thanks{E-mail: camero@ice.cat}, N. Rea$^{1}$, N. Bucciantini$^{2,3}$, M. A. McLaughlin$^{4}$, P. Slane$^{5}$,  \newauthor
B. M. Gaensler$^{6}$, D. F. Torres$^{1,7}$, L. Stella$^{8}$, E. de O\~na$^{1}$, G. L. Israel$^{8}$, F. Camilo$^{9,10}$ \newauthor
and A. Possenti$^{11}$
\vspace*{0.09cm}\\
$^{1}$  Institut de Ci\`{e}ncies de l'Espai, (IEEC-CSIC), Campus UAB, Fac. de Ci\`{e}ncies, Torre C5, parell, 2a planta, 08193 Barcelona, Spain\\
$^{2}$  INAF - Osservatorio Astrofisico di Arcetri, L.go E.~Fermi 5, 50125, Firenze, Italy\\
$^{3}$  INFN - Sezione di Firenze, Via G.~Sansone 1, 50019 Sesto Fiorentino, Firenze, Italy\\
$^{4}$  Department of Physics, West Virginia University, Morgantown, WV 26501, USA\\
$^{5}$  Harvard-Smithsonian Center for Astrophysics, 60 Garden St. Cambridge, MA 02138, USA\\
$^{6}$  The University of Sudney, Room 216, 44 Rosehill Street, Redfern, NSW 2016, Australia\\
$^{7}$  Instituci\'{o} Catalana de Recerca i Estudis Avancats (ICREA)\\
$^{8}$  INAF - Osservatorio Astronomico di Roma, Via Frascati 33, I-00040 Monteporzio Catone (Roma), Italy\\
$^{9}$  Columbia Astrophysics Lab, Columbia University, New York, NY 10027, USA\\
$^{10}$ Arecibo Observatory, HC3 Box 53995, Arecibo, PR 00612, USA\\
$^{11}$ INAF-Osservatorio Astronomico di Cagliari, loc. Poggio dei Pini, strada 54, 09012 Capoterra, Italy}

\input psfig.sty

\begin{document}

\pagerange{\pageref{firstpage}--\pageref{lastpage}} \pubyear{2012}

\maketitle

\label{firstpage}

\begin{abstract}

We present new imaging and spectral analysis of the recently discovered extended X--ray emission around the high magnetic field rotating radio transient RRAT\,J1819--1458.  We used two {\it Chandra} observations performed for this object in 2008 May 31 and 2011 May 28, respectively. The diffuse X--ray emission was detected with a significance  of $\sim$19$\sigma$ in  the image obtained by combining the two observations. Neither long-term spectral nor timing variability have been observed from the source or the nebula. RRAT\,J1819--1458 shows an unusual high X--ray efficiency of $\eta_{X} \equiv L_{X(0.3-5\,keV)}/\dot{E}_{rot} \sim$0.15 at converting spin-down power into  X--ray  luminosity.  The most favourable scenario for the origin of this extended X--ray emission is either a pulsar-wind nebula (PWN) or a scattering halo. A magnetically powered scenario for the extended emission is viable only in the case of a Compton nebula, while can be tentatively disfavoured in the case of synchrotron emission.

\end{abstract}



\begin{figure*}
\hbox{
\psfig{figure=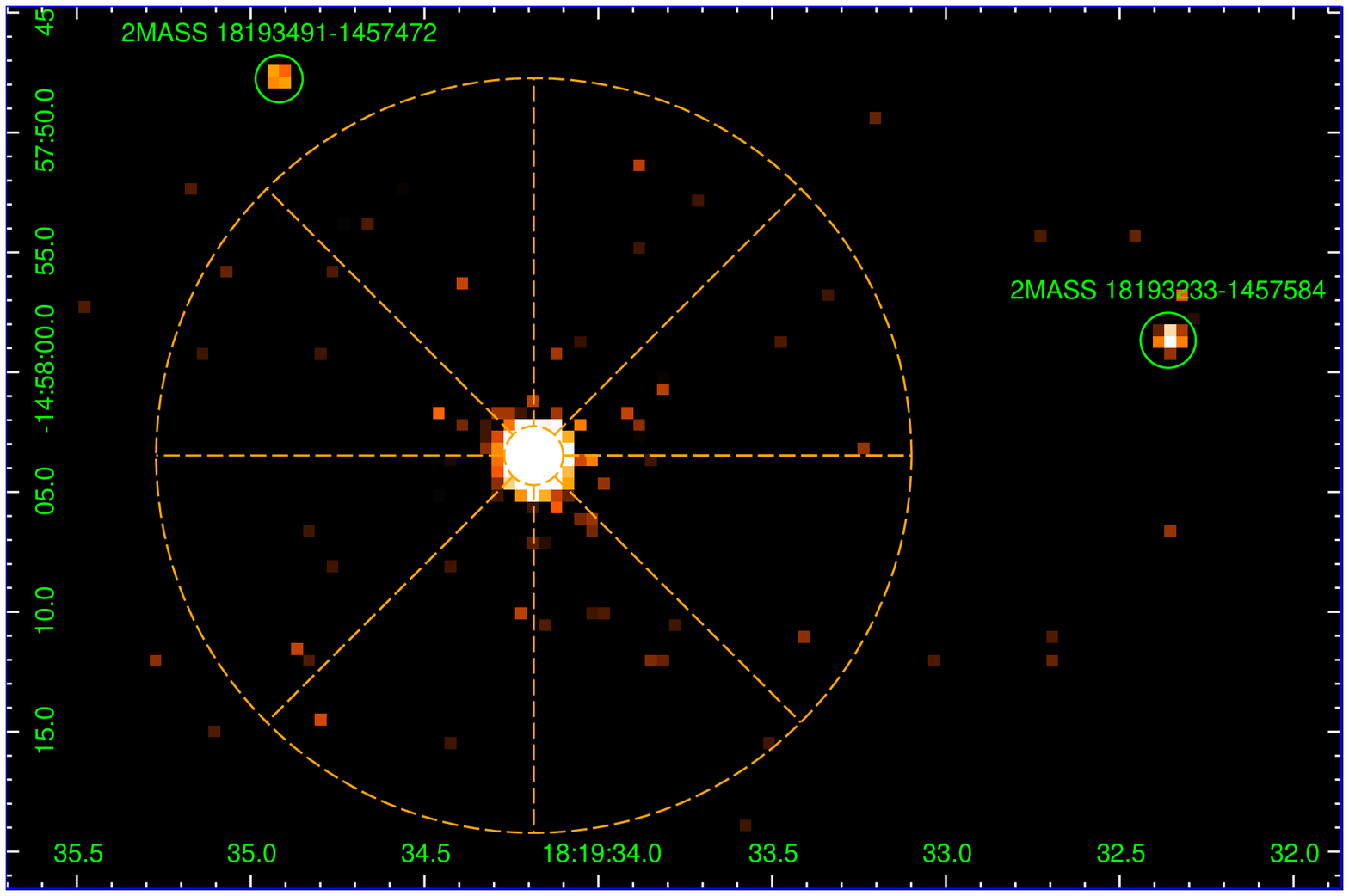,width=8.75cm,height=6cm}
\hspace{-0.05cm}
\psfig{figure=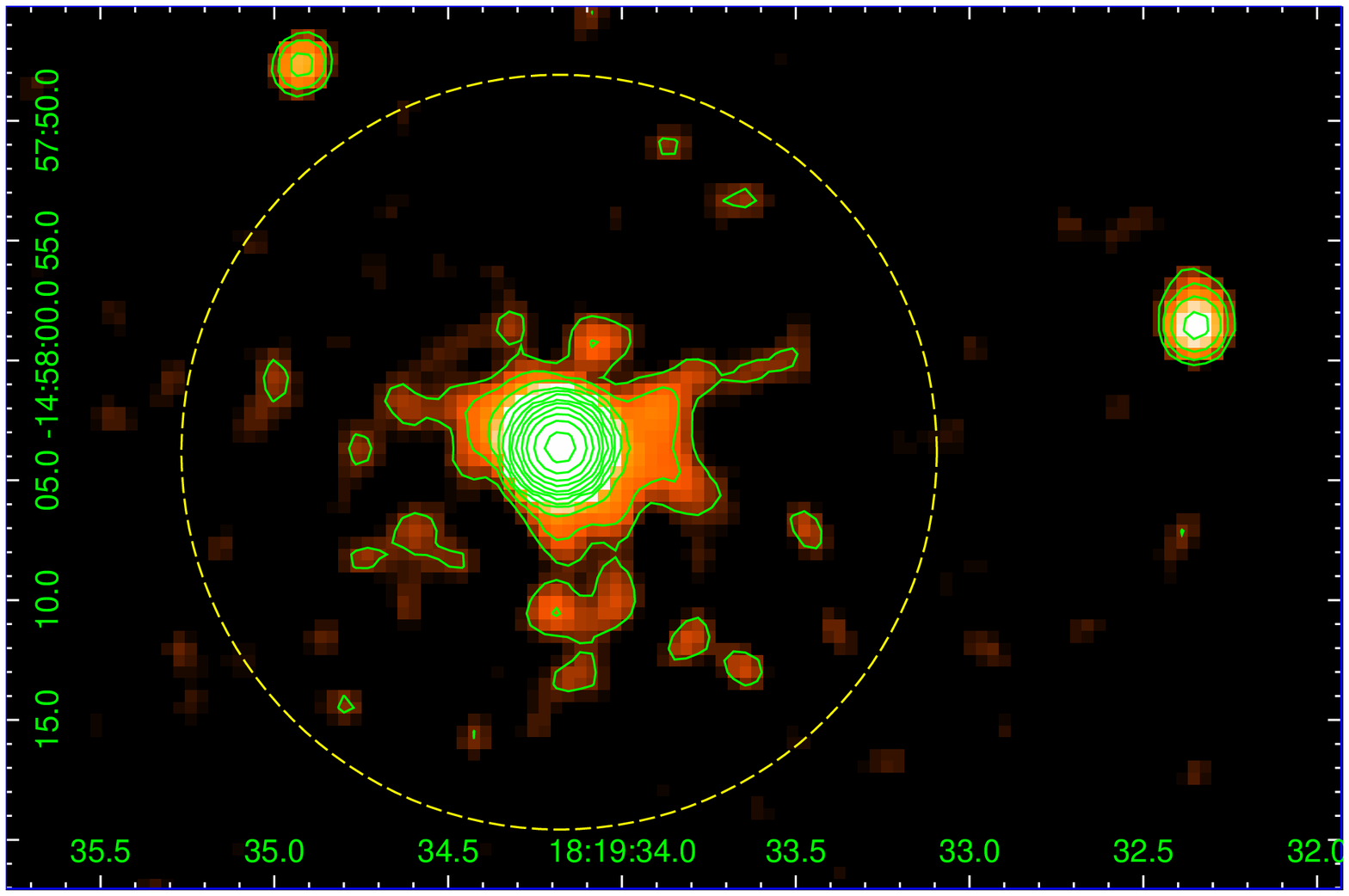,width=8.75cm,height=6cm}
}
\vspace*{0.2cm}
\caption{{\em Left panel}: Combined 0.3--10\,keV log image of RRAT J1819--1458, using both \textit{Chandra} ACIS-S observations from 2008 and 2011, with a panda region of outer radius 16$\arcsec$ and inner radius 5$\arcsec$ overplotted. The offsets from the RRAT J1819--1458 position are $\sim$19$\arcsec$ and $\sim$27$\arcsec$ for the 2MASS 18193491--1457472 and 18193233--1457584 sources, respectively.
 {\em Right panel}. Smoothed image using a Gaussian function with a radius of 3 pixels and with 3$\sigma$ contours of the extended emission and a circular region of 16$\arcsec$  overplotted. Colors are proportional to the log of the X--ray intensity. North is up, and east is to the left. One ACIS-S pixel corresponds to 0$\arcsec$.492.}
\label{ima}
\label{sf}
\end{figure*}

\section{INTRODUCTION}

Rotating Radio Transients (RRATs) are radio pulsars that were discovered through their sporadic radio bursts \citep{mcLaughlin06}.   There are $\sim$70 currently known RRATs,  with spin periods ranging from 0.1 to 7\,s \citep[and references therein]{keane11}. 

At a radio frequency of 1.4 GHz, radio bursts are observed from \f ~roughly every $\sim$3 minutes with the Parkes telescope. The spin period of RRAT\,J1819--1458  is 4.3\,s ($\dot{P}\sim$ 3.2$\times$10$^{-13}$\,s\,s$^{-1}$; Lyne et al. 2009), with  a characteristic age of 117\,kyr and a dipolar magnetic field of  $B\sim$5$\times$10$^{13}$\,G. The spin-down energy loss rate measured for this source is  \.{E}$_{rot}\sim$3$\times$10$^{32}$ erg s$^{-1}$.  Two glitches have been detected, with one of these showing anomalous post-glitch recovery that suggests \f ~originated in the magnetar region of the period-period derivative diagram \citep{lyne09}.

RRAT\,J1819--1458   is the only source of this type also detected at  X--rays energies so far \citep{reynolds06,mcLaughlin07,rea_cLaughlin08,kaplan09}. The distance inferred from its dispersion measure (196$\pm$3 pc\,cm$^{-3}$)  is 3.6\,kpc, with at least a 25$\%$ of uncertainty \citep[][and references therein]{mcLaughlin07}. The spectrum of \f ~is well modeled by an absorbed blackbody (N$_{H} \sim 6 \times10^{21}$~cm$^{-2}$ and kT$ \sim 0.14$\,keV) with a broad spectral absorption line at $\sim$1~keV  (McLaughlin et al. 2007, Rea et al. 2009). The X--ray luminosity is  $L_{\rm X}\sim$4$\times$10$^{33}$(d/3.6 kpc)$^2$ erg s$^{-1}$, more than one order of magnitude higher than the spin-down luminosity.

Diffuse X--ray emission was found around RRAT J1819--1458 in a $\sim$30\,ks \textit{Chandra} observation carried out in 2008, with a luminosity of $\sim$10$^{32}$ erg s$^{-1}$ and extending to $\sim$13$\arcsec$ from the source \citep{rea09}. Furthermore, the pulsar's error circle was refined performing a boresite correction of this new \textit{Chandra}  observation, using a 2MASS field star present in both X--ray and infrared images. This resulted in  an accurate position of right ascension $\alpha = $18$^h$19$^m$34$^s$.173 and declination $\delta = $-- 14$^\circ$58$\arcmin$03$\arcsec$.57 (J2000; with a 1$\sigma$ error of $\sim$0.3$\arcsec$ in both coordinates; Rea et al.\ 2009). 

So far there is no evidence of an optical counterpart for \f ~\citep[see e.g.][]{dhillon11}. However, near-infrared  observations (J, H and Ks filters) resulted in the identification of a possible candidate near-infrared counterpart which  is the only source within the 1$\sigma$ X--ray positional error circle \citep{rea10}. 

In this work, we present the study resulting from the reduction and combined analysis of two {\it Chandra} observations for RRAT J1819--1458, performed on 2008 May 31 (Rea et al. 2009) and 2011 May 28 (this work). Observations and data reduction are reported in Section 2, the analysis and results in Section 3, and a discussion in Section 4.

\section{OBSERVATIONS AND DATA REDUCTION}

The \textit{Chandra} X--ray Observatory observed RRAT J1819--1458  with the Advanced CCD Imaging Spectrometer (ACIS) instrument on 2008 May 31 (ObsID 7645) for 30\,ks  and again in 2011 May 28 (ObsID 12670) for 90\,ks,  both in VERY FAINT (VF) timed exposure imaging mode. 

For both observations, we used a 1/8 subarray, which provides a time resolution of 0.4 s, and the typical ACIS-S imaging and spectral configurations. The source was positioned in the back-illuminated ACIS-S3 CCD at the nominal target position. Standard processing of the data was performed by the \textit{Chandra} X--ray Center to Level 1 and Level 2 (processing software DS 7.6.11.6 for ObsID 7645 and 8.4.3 for ObsID 12670). 

 In this work we have used CIAO software (ver.\,4.4) for the posterior processing and analysis of the data, resulting in a final exposure time of 27.88~ks for the first observation and 80.40~ks for the second one.

\section{ANALYSIS AND RESULTS}

\subsection{Imaging}

To study the extended X--ray emission found by \citep{rea09} in more detail,  we proceeded with the extraction of a combined image in the 0.3--10 keV energy range, using the two  \textit{Chandra} observations described in the previous Section. Before merging the two observations, we reprojected the events onto the same tangent plane.  Therefore, for aligning and merging event files from different ObsIDs we used the CIAO tool \texttt{reproject$\_$image}.   Figure~\ref{ima} shows the resultant combined image where  diffuse extended X--ray emission is clearly visible around the compact object. 
 
We applied the CIAO \texttt{wavdetect} tool  to the $\sim$90\,ks ACIS-S cleaned image and found   RRAT J1819--1458 at $\alpha  = 18^{\rm h}19^{\rm m}34.178^{\rm s}$ and  $\delta = -14^\circ58\arcmin03\arcsec.662$ (J2000), with a statistical error of 0$\arcsec$.007 radius, and 2  X--ray bright stars in the field detected at a significance of $>$5$\sigma$. The second star detected is the Two Micron All Sky Survey (2MASS)\footnote{http://www.ipac.caltech.edu/2mass} star 18193233--1457584 at  $\alpha = 18^{\rm h}19^{\rm m}32.34^{\rm s}$ and $\delta = -14^\circ57\arcmin58\arcsec.68$ (J2000), with an accuracy of 0$\arcsec$.07 radius (catalog position: $\alpha = 18^{\rm h}19^{\rm m}32.34^{\rm s}$ and $\delta = -14^\circ57\arcmin58\arcsec.41$, with an accuracy of 0$\arcsec$.08 radius \citep[see also][]{rea09}. 

The third source detected, at  $\alpha = 18^{\rm h}19^{\rm m}34.92^{\rm s}$ and  $\delta = -14^\circ57\arcmin47\arcsec.68$ (statistical error of 0$\arcsec$.12 radius),  was consistent with the 2MASS star 18193491--1457472  (catalog position of  $\alpha = 18^{\rm h}19^{\rm m}34.912^{\rm s}$ and $\delta = -14^\circ57\arcmin47\arcsec.22$, with a 0$\arcsec$.11 error radius). We then proceeded  to perform a bore-site correction of the field to refine the RRAT\,J1819–-1458 position and error circle. Assuming a physical association between the 2MASS stars and their coincident X--ray sources, the final  RRAT\,J1819–-1458 position is $\alpha  = 18^{\rm h}19^{\rm m}34.18^{\rm s}$ and  $\delta$ = --14$^{\circ}$58$\arcmin$03$\arcsec$.7,  with a 1$\sigma$ associated error circle of 0$\arcsec$.2 radius (computed doing a quadratic mean of all the positional and statistical errors plus the 2MASS catalog intrinsic systematic errors). \\


\begin{table*}
\begin{center}
\caption{ Best-Fit  Spectral Parameters$^{a}$.}
\begin{tabular}{llllllllll}
  \hline\noalign{\smallskip}
 \hline\noalign{\smallskip}

                                 &\hspace{0.3cm} RRAT&\hspace{-0.3cm}J1819--1458&   &&&  \hspace{-0.3cm} DIFFUSE  &\hspace{-0.4cm}  EMISSION&&\\
  \noalign{\smallskip}                               
   ObsID            & \textit{N$_{\rm H}^b$}    &   \textit{T$_{BBody}^{c}$}  &   \textit{E$_{gauss}^{c}$}      & Flux$^e$                         &\hspace{-0.15cm}    $\chi_r^2$(DOF)             & \textit{N$_{\rm H}^b$}     &  \textit{$\alpha$}                    & Flux$^e$    &   $\chi_r^2$(DOF)  \\
                                 &                            & \textit{norm$^d$}        & \textit{$\sigma^{c}$}              &                        & \hspace{-0.14cm}                                           &                                & \textit{norm$^f$}                        &                    &            \\
  \hline \noalign{\smallskip} \hline \noalign{\smallskip}

    7645                    &  0.6 (fixed)        &   0.129$\pm$0.003          &   1.13$\pm$0.04             &  1.37 $\pm$0.05  & \hspace{-0.14cm} 1.05(29)          & $<$0.7                             & 3.6$\pm$0.4                      &  0.22$\pm$0.04   & 1.01(6)  \\
   (May 2008)          &                            &  1600${+300\atop-200}$  &    0.14$\pm$0.05        &                              &\hspace{-0.14cm}                        &                                          &  10.3$\pm$0.2  &                              &              \\

  \hline\noalign{\smallskip}

   12670                   & 0.6 (fixed)        &  0.129$\pm$0.002             &  1.16$\pm$0.03         &  1.30$\pm$0.02   &\hspace{-0.14cm}   1.30(44)              &    $<$0.7                            &  3.5$\pm$0.3                      &   0.25$\pm$0.04    &  1.10(17)       \\
    (May 2011)           &                       &  1400${+200\atop -100}$ &  0.17$\pm$0.04         &                             &\hspace{-0.14cm}                             &                                            &  9.7$\pm$0.8                       &                               &           \\
  \hline\noalign{\smallskip}
  
  7645$+$12670    &   0.6 (fixed)       & 0.130$\pm$0.002             & 1.16$\pm$0.03          & 1.35$\pm$0.02   & \hspace{-0.14cm}  1.10(45)                &     $<$0.9                        &  3.7$\pm$0.3           &  0.23$\pm$0.02    &   1.26(19)         \\
         (combined)    &                         &  1500$\pm$100  &  0.17$\pm$0.03        &                           & \hspace{-0.14cm}                              &                                          &  8.7$\pm$0.6                     &                              &          \\

  \hline\noalign{\smallskip} 

   \end{tabular}
 \end{center}
$^{a}$Results of the spectral modeling with a {\tt phabs*bbodyrad*gabs} and {\tt phabs*power}, for source and diffuse emission, respectively; \\ $^{b}$ All errors are at 90$\%$ confidence level; N$_{\rm H}$ in units of  $10^{22}$\,cm$^{-2}$; \\ $^{c}$  Gaussian absorption line  energy and width in keV units; \\ $^{d}$ a constant of value R$_{km}^2$/d$_{10}^2$, where R$_{km}$ is the source radius (km)  and d$_{10}$ is the distance  in units of 10 kpc; \\ $^{e}$ absorbed flux  in units of $\times 10^{-13}$\,erg\,cm$^{-2}$\,s$^{-1}$   (0.3--5\,keV); \\ $^f$ Photon Index with normalization units in 10$^{-6}$ photons  keV$^{-1}$cm$^{-2}$s$^{-1}$ at 1 keV.\\
 
\label{specfits}
\end{table*}

\subsection{Timing}

For the timing analysis, we first referred the arrival time of each photon to the barycenter of the solar system using the CIAO tool  \texttt{axbary}. Then, we used the \texttt{dmextract}  tool to create background-subtracted lightcurves, using the time resolution of the data ($\sim$0.4\,s).  For this, we extracted the source photons on each individual observation from a circular region with 2$\arcsec$.5 radius, and another one for the  background,  far from the source.  

Using the \texttt{Xronos} package, we folded both  X--ray data sets  using the radio ephemeris \citep{lyne09}, confirming the sinusoidal X--ray modulation found with \textit{XMM-Newton} \citep{mcLaughlin07} and \textit{Chandra} \citep{rea09}. In addition, we performed a periodicity search obtaining a P$_{spin}$=4.26328(6)\,s (epoch= MJD 54\,617),
in agreement with previous work. The  0.3--5 keV pulsed fraction for the new \textit{Chandra} observation (ObsID 12670) was of 
 31$\pm$4$\%$, defined as ($F_{max}$ -- $F_{min}$)/($F_{max}$ + $F_{min}$),  with $F_{max}$ and $F_{min}$ the maximum and minimum counts of the background-corrected X--ray pulse profile. The shape of the pulse profile, and the pulsed fraction, are consistent with past measurements \citep{reynolds06,mcLaughlin07,rea09},  showing no  evidence for long-term variability.

\subsection{Spectroscopy}

\begin{figure}
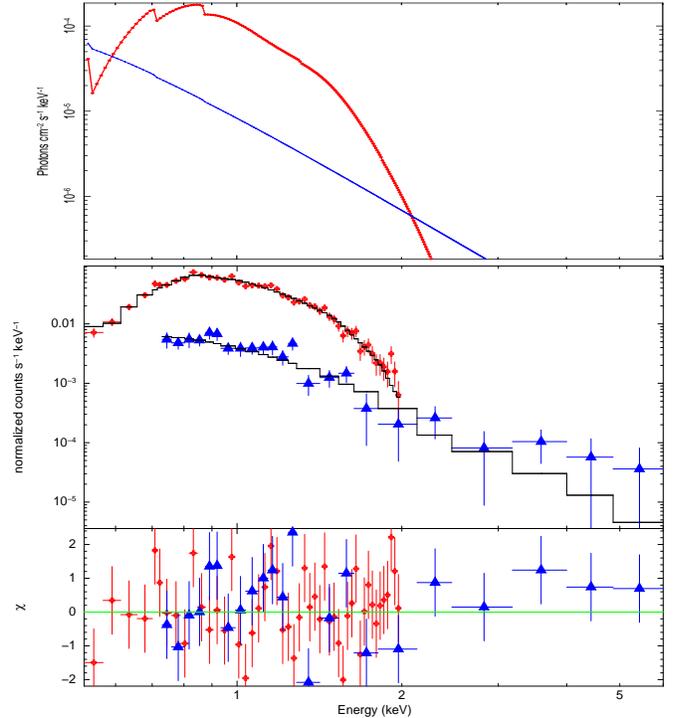

\hspace*{0.9cm}\psfig{file=Figurita3.ps,width=5.75cm,height=5cm}\vspace*{-2.48cm}
\hspace*{0.9cm}\psfig{figure=Figurita4.ps,width=6.162cm,height=8.5cm}
\caption{{Best-fit deconvolved model (top panel) for RRAT J1819--1458 (red circles) and the extended X--ray emission (blue triangles).  The spectrum of RRAT J1819--1458 was modeled with an absorbed blackbody  plus a $\sim$1\,keV absorption line,  and an absorbed power-law  for the extended X--ray emission. The normalized spectra and residuals are also shown in the middle and bottom panels, respectively}. 
}
\label{sp}
\end{figure}

\subsubsection{\f}

We used the \texttt{specextract} script, which uses a combination of CIAO tools, to extract source and background spectra for a point-like ACIS source like RRAT\,J1819--1458. To extract only the photons from the point source for both observations,  a circular region with 2$\arcsec$.5 radius and a circular background region of radii 18$\arcsec$ (far from the source) were used. We neglected  in the source spectral analysis the projected emission from the extended X--ray nebula since it only contributed  $\sim$3$\%$ of the counts. The point source spectrum was rebinned so as to have at 
least 25 counts per spectral bin, so that $\chi^2$ statistics could be used.

We modeled each spectrum using the XSPEC v.12.7.0u analysis package. Following previous studies \citep{mcLaughlin07,rea09}, we fit the continuum of each spectrum with an absorbed blackbody.  However, a single blackbody fit does not represent the spectrum properly ($\chi_r^2\sim$1.7; 47 dof). We added to that model  the previously detected absorption line at 1~keV \citep{mcLaughlin07,rea09}, which we modeled with a Gaussian function ({\tt phabs*bbodyrad*gabs}  in XSPEC notation). Table~\ref{specfits}  shows the spectral parameters obtained for the best fit.  Since the Hydrogen absorption column was not found to significantly vary between observations, we fixed this parameter in both observations at the value found by McLaughlin et al. (2007) using an {\em XMM-Newton} observation.  This allowed us to better constrain the 1\,keV feature with respect to the previous {\em Chandra} observation, although it is not as well resolved as with the \textit{XMM-Newton} observation (Miller et al. 2012 in preparation), due to the different effective area of the two satellites. The inferred blackbody radius is 14$\pm$8\,km, assuming a 3.6 kpc distance. 

In order to increase the signal to noise of the spectrum we proceeded to combine the spectra created for ObsId 7645 and 12670  and associated responses. For this we used the CIAO tool \texttt{combine$\_$spectra} and then we rebinned the final $\sim$0.5--2\,keV combined spectrum  to have at least 25 counts per spectral bin (see Figure~\ref{sp}). We used in addition the FTOOL \texttt{mathpha}, yielding similar results. Once again, the combined spectrum was modeled in XSPEC with the same model as before (see Table~\ref{specfits}).  Fitting the two individual spectra simultaneously with the same model also yielded similar results.

In the new \textit{Chandra} observation (ObsID 12670), the point source RRAT\,J1819--1458 has an ACIS-S 0.3--10 keV count rate of 0.0374(7) counts s$^{-1}$ (background subtracted) and in the combined observation the count rate is 0.0391(6) counts s$^{-1}$ (a total of $\sim$4225 counts).

\begin{figure*}
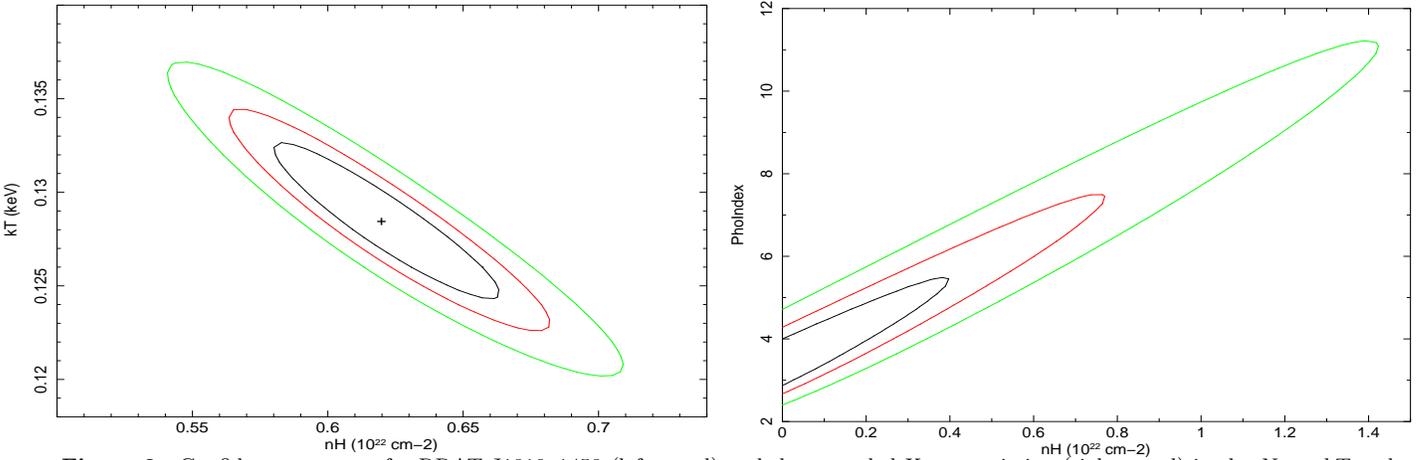

\begin{center}
\psfig{figure=Figurita5.ps,width=7cm,height=8cm} 
\vspace*{-8cm} 
\hspace*{+8cm}
\psfig{file=Figurita6.ps,width=7cm,height=8cm}
\caption{ Confidence contours for RRAT J1819--1458 (left panel) and the extended X--ray emission (right panel) in the $N_H-kT$ and $N_H-Gamma$ spaces, respectively. From inside to outside  1$\sigma$, 2$\sigma$, and 3$\sigma$ contours are displayed.  }
\end{center}
\label{cont}
\end{figure*}

\subsubsection{The diffuse  X--ray emission}

To extract the photons from the diffuse X--ray emission for both observations,  we followed the same procedure as in \cite{rea09} and selected an annular region of  inner radius 2$\arcsec$.5  and outer radius of 20$\arcsec$, to ensure that the whole extended X--ray emission was included (see Fig.~\ref{ima} and  Sect.~3.4). For the background we selected a similar annular region but  far from the extended X--ray emission. An absorbed power law provides a good fit to the data (see Table~\ref{specfits}; the spectrum was grouped with  at least 25 counts per bin), but the spectral parameters are somewhat poorly determined due to the small number of counts \citep[see also][]{rea09}. 

Following the same approach as for \f , we then create a combined spectrum using combine$\_$spectra (see previous Section).  Figure~\ref{sp} shows the obtained combined $\sim$0.8--7\,keV spectrum for the extended source (the spectrum was also grouped with at least 25 counts per bin). The background contribution is $\sim$35$\%$ for the extended emission. The extended X--ray emission has a 0.3--10 keV count rate of 0.0077(5) counts s$^{-1}$ (background subtracted) and a total of $\sim$830 counts. In Table~\ref{specfits} we display the spectral parameters resulted from the best fitting for all the spectra. We note that fixing the Hydrogen absorption column to the pulsar's value showed systematic departures from the data at high energies, preventing us from constraining  the rest of the parameters.  This is an effect of poor statistics,  with the limits of the N$_H$ derived for the nebula being consistent with the value obtained for the point source (see Fig. \ref{cont}).

 \subsection{The diffuse X--ray emission structure}

To infer the significance and estimate the luminosity of the whole diffuse emission in the combined image, we built the combined Chart/MARX point-spread function (PSF). To do this, we first  built a Chart/MARX PSF for each individual observation, using both the RRAT J1819--1458 spectrum and its corresponding exposure time.  The CIAO tool \texttt{reproject$\_$image} then reprojected the events onto the same tangent plane, and created a final combined PSF image. 

In Figure~\ref{sf}, we compare the surface brightness radial distribution of the combined Chandra observation of RRAT\,J1819--1458~with that of the combined Chart/MARX PSF plus a background level. Both surface brightnesses were obtained by extracting counts from 50 annular regions (each 2 pixels wide) centered  on the source position, and for the RRAT J1819--1458 one, after removal of the serendipitous point sources in the field. This figure shows that the extended emission becomes detectable around 5 pixels ($\sim$2$\arcsec$.5) from the peak of the source PSF.

To compute the significance of the diffuse X--ray emission around RRAT\,J1819--1458, from the combined image we extracted all the photons from an annular region of 2$"$.5--20$"$ radii, and we subtracted from it the background extracted from a similar region far from the source (but in the same S3 CCD).  This resulted in an excess of 790$\pm$18  counts, which corresponds to a detection significance of  $\sim$19$\sigma$.

Finally, we studied any possible change in the morphology of the extended X--ray emission between the 2008 and 2011 observations. To account for this we applied the Kolmogorov-Smirnov  statistic (KSTWO), using the count-rate/pixel$^{2}$ from each surface brightness radial distributions as input vectors, to study whether the two images were drawn from the same distribution.  The associated probability resulted to be $\sim$0.7, meaning that the two data sets most likely come from the same distribution. In addition, we did not find in the combined image any azimuthal asymmetry in the diffuse emission (see right panel of Fig. \ref{sf}).

\begin{figure*}
\begin{center}
\psfig{file=Figurita7.eps,width=8.36cm,height=7cm}
\vspace*{-7.2cm} 
\hspace*{+8.5cm}
\hspace*{-0.35cm}\psfig{file=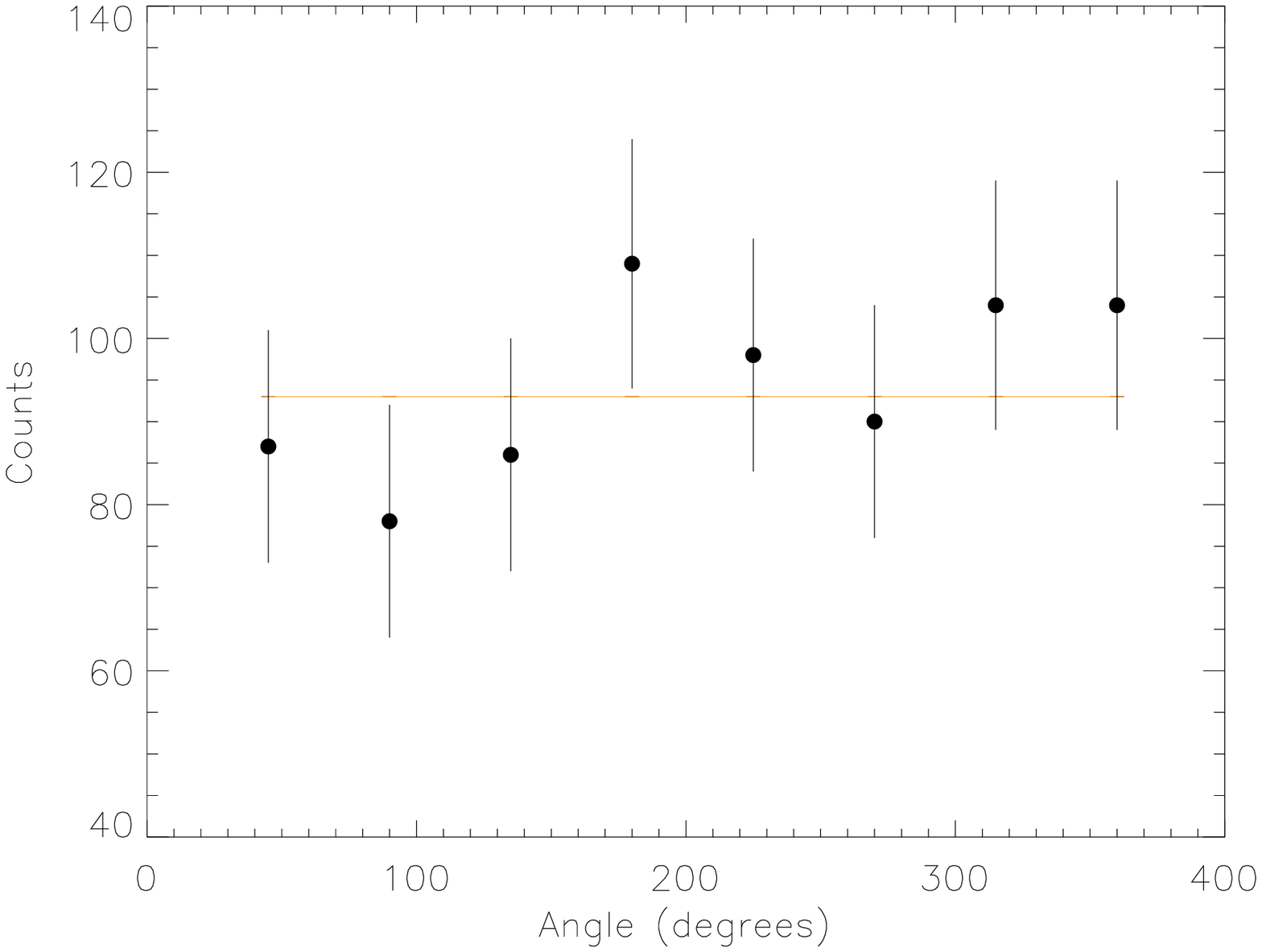,width=9.5cm,height=6.6cm}
\caption{ \textit{Left.} Surface brightness of the background-subtracted ACIS-S image of RRAT J1819$-$1458 (red open squares) and of the Chart/MARX PSF plus a constant background (blue circles). \textit{Right}. Azimuthal distribution of counts (background subtracted) in the extended X--ray emission for a panda region divided in 8 sections (see  left panel of Fig.~\ref{ima}; West, to the right, corresponds to 0 degrees, with the angle increasing counterclockwise). A constant fit is overplotted (reduced $\chi^2\sim$0.7; constant $\sim$94 counts).}
\label{}
\end{center}
\end{figure*}

\section{Discussion}

{In the present work we have not found long-term variability in spectral and timing X--ray properties  for RRAT\,J1819--1458. The spectral continuum for the point source was well fitted with an absorbed blackbody model, in good agreement with previous results  \citep{mcLaughlin07,rea09}. The  previously reported absorption line at 1~keV \citep{mcLaughlin07,rea09} is  visible in all the observations. For the extended X--ray emission the spectral parameters did not change in time, and are compatible with the result from \cite{rea09}.  In addition, the diffuse X--ray emission was detected in the combined image with a significance  of $\sim$19$\sigma$, substantially improving the detection level reported in its discovery \citep{rea09}.

The energies of pulsar wind electrons and positrons range from $\sim$1 GeV to $\sim$1 PeV, placing their synchrotron and inverse
Compton  emission into radio--X--ray and GeV--TeV bands, respectively. This multi-wavelength
emission can be seen as a pulsar-wind nebula \citep[and references therein]{kargaltsev12}. 
To date, the exact physical origin and acceleration mechanism of the high-energy particles in the pulsar  winds are poorly understood, and not all nebulae can be easily explained as spin-down-powered PWNe.

In \citet{rea09} we discussed different scenarios for the origin of the extended emission detected around RRAT\,J1819--1458.  One option was that the  extended emission we observe is part of the remnant of the supernova explosion which formed RRAT\,J1819--1458, unlikely for an object of 117 kyr. Another possibility  was a bow-shock nebula due to the pulsar moving supersonically through the ambient medium, but this was ruled out since the projected velocity in the case of a bow shock would be rather small (v$_p\sim$20\,km\,s$^{-1}$; see Rea et al. 2009 and references therein). 

Other possibilities are that \f ~could power a sort of PWN, or the extended X--ray emission around the pulsar might be explained as a magnetic nebula,  or as a scattering halo as  for 1E\,1547--5408 \citep{vink_bamba09,olausen11} and Swift\,J1834.9-0846 \citep{younes12,esposito12}.

In the following we will investigate the PWN hypothesis for the extended emission found in coincidence with RRAT~J1819$-$1548. This is mostly a qualitative analysis, trying to constrain a possible PWN interpretation. A distance of $3.6$ kpc is assumed. At that distance the observed diffuse emission extending to  $20 \arcsec$ corresponds to a radius of the nebula: $R_{\rm pwn} \approx 1$\,ly.  X--ray PWNe are usually observed in coincidence with young pulsars ($\tau \sim 10^{3-4}$ yr), which have high spin-down luminosity to power the observed emission. Their X--ray luminosity  in the 2--10\,keV band is found to positively correlate  \citep{Xiang-Hua-Li08} with the spin-down power $\dot{E}$ (more energetic pulsars tend to inject more high energy particles).  The X--ray luminosity is also found to anticorrelate with the characteristic age $\tau$ (older systems tend to be less compact and with a lower magnetic field). However, we caveat that in the latter correlations there could be important selection effects, as well as a scatter by several orders-of-magnitude.

For the nebula associated with RRAT~J1819$-$1548, extrapolating the observed flux and using the observed photon index, we find  $L_{\rm X}$[2-10 keV]$\sim$ 1.5$\times$10$^{30}$ erg s$^{-1}$ $\pm$ 50$\%$, while the correlations found by \citet{Xiang-Hua-Li08}, with $\dot{E} \sim 3\times10^{32}\,{\rm erg\; s}^{-1}$  would result in  $L_{\rm X} =3.1\times10^{27}{\rm erg\; s}^{-1} $. It is evident that compared with other pulsars showing PWNe, this nebula show a relatively high efficiency in converting rotational energy in X--ray luminosity.

However the relations in \cite{Xiang-Hua-Li08}, derived for young and energetic pulsars, when extrapolated to  low values of $\dot{E}$, typical for  RRAT~J1819$-$1548, have large uncertainties.  Thus the inferred value for $L_{\rm X}$  with respect to our measured value, despite a few orders of magnitude difference, is compatible within a couple of standard deviations of the model predictions \citep{Xiang-Hua-Li08}. 

Another peculiar aspect of the detected extended X--ray emission is the very steep photon index $\Gamma_{\rm pwn} $. If one assumes an injection spectrum of X--ray emitting particles $\dot{N}(E)\propto E^{-p}$ one gets for the photon index: $(p+1)/ 2 < \Gamma_{\rm pwn} < (p+2)/2$ \citep{pacini-salvati}. In the following discussion we will assume that the synchrotron cooling time of X--ray emitting particles  is shorter than the age of the nebula.
Synchrotron cooling timescale for X--ray emitting particles in magnetic fields of $\sim 10 \mu$G (see the discussion below on the possible values of the magnetic field), are much smaller than the spin-down age of the system. Therefore, an injection spectrum with $p=5$ is needed to explain the observed photon index, while typical values in other PWNe are found to be $ p \sim2.1-2.5$. RRAT~J1819$-$1548,  with $L_{\rm X}$[0.3--5\,keV] = $3.6\times10^{31}{\rm erg\; s}^{-1} \sim0.12\times  \dot{E}$,  would have to inject  $>12$\% of the total spin-down power  in the form of X--ray emitting particles.
Given the steep photon spectrum that we observe, the X--ray efficiency (12\%) with respect to  $\dot{E}$ is insensitive to our choice for the high energy limit of the energy band (5\,keV). However it is very sensitive to the low energy limit  (0.3\,keV). This suggests that the observed photon spectrum cannot extend smoothly below this energy, otherwise the total efficiency  could rapidly exceed 100\%. We will assume in the following discussion that a spectral break is present at 0.3\,keV.

Let us now attempt to build a standard PWN model, and see if, and under which conditions (limits) the observed properties can be reproduced. We want to stress here, that given the paucity of data (only an X--ray flux and photon index are available) and the uncertainties in our assumptions (for example the distance), this is mostly an attempt to constrain the plausibility of a PWN interpretation.
In general the pulsar wind will inject into the nebula both relativistic particles and magnetic field: $\dot{E} = \dot{E}_{\rm part} + \dot{E}_{\rm mag}$  (with $\dot{E}_{\rm part}$ being the energy injected in particles and $ \dot{E}_{\rm mag}$ the energy injected in magnetic field).  It is found that, in many PWNe, a single power-law injection spectrum for the particles cannot reproduce the observed inegrated specta from radio to X--rays. To fit the integrated broad band spectrum, one requires the injected particles to have at least an energy distribution in the form of a broken power-law \citep{bucciantini_al11}.  If we call $E_b$ the break energy, then the injection spectrum of the particles is: \\
\begin{eqnarray}
\dot{N}(E < E_b) \propto (E/E_b)^{-\alpha_1},\\
\dot{N}(E > E_b) \propto (E/E_b)^{-\alpha_2}.
\end{eqnarray}

From the observed X--ray photon index, we can fix the high energy slope of the injected spectrum $\alpha_2 = 5$. For the low energy part one can adopt  the flattest value measured in PWNe in radio, $\alpha_1 = 1$. We stress here that for the following discussion, this is the most optimistic assumption, that minimizes the amount of energy injected below the break $E_b$. The energy injected in particles is:
\begin{equation}
 \dot{E}_{\rm part} = \int_0^\infty E\dot{N}(E)dE. 
\end{equation}
On the other hand, the energy injected above the break  is
\begin{equation}
 \dot{E}_{\rm part}(E>E_b) = \int_{E_b}^\infty E\dot{N}(E)dE \approx 0.25~\dot{E}_{\rm part}.
\end{equation}
Thus of all the energy injected in particles, for our values  of $\alpha_1 $ and $\alpha_2$, $\sim25\%$ is in particles with E $>$ E$_b$ (above the break). Of course this number can be further increased if one introduces a low energy cutoff, which however is not observed in other PWNe, or if a smooth break is assumed.  Taking into account the synchrotron cooling  and the inferred injection spectrum at high energy, one can show that in order to have a $\sim$12\% efficiency in the radiated X--rays, as observed, one needs to inject $\sim 25-30$\%  of the total spin-down power  in the form of X--ray emitting particles.

We will call $E[\nu_X]$  the energy of particles responsible for the emission photons with energy $\nu_X$. For the synchrotron emission the value of this energy is a function of the energy of the emitted photon and of the magnetic field. Despite the fact that the magnetic field is not known, we will show that information on the inferred injection spectrum, can be used to constrain it. Taking into account all that was stated before, we  conclude that: 
\begin{itemize}
\item  $E_b$ cannot be $\gg E[0.3{\rm keV}]$, otherwise we would get a harder spectrum than the one observed.
\item $E_b$ cannot be $\ll E[0.3{\rm keV}]$, otherwise the fraction of energy injected in X--ray emitting particles would be $<$25$\%$ of $\dot{E}_{part}$ and $<$ 25$\%$ $\dot{E}$ (and we could not explain the high X--ray efficiency). 
\end{itemize}
Using the above constraints we can put an upper limit on the value of the nebular magnetic field. Assuming $E_b \approx E[0.3{\rm keV}]$, and recalling that 25\% of the energy must be injected above the break, we find  that almost ~75\% of the total spin-down energy is injected in particles below the break.  This implies that $\dot{E}_{\rm mag} \ll \dot{E}$, i.e. the high X--ray efficiency requires that most of the energy is injected into particles, with little left to be injected into the magnetic field. The magnetic energy content of the nebula is $ E_{\rm mag}\sim \dot{E}_{\rm mag}\times\tau \sim \eta_B \times \dot{E}\times\tau$, where $\eta_B$ can be roughly assumed $ < 0.1$, which gives a magnetic field  $B_{\rm pwn} < 20 \mu$G.  On the other hand, we can show that $B$ must be greater than $\sim$1$\mu$G. In fact if $B$ was  $\leq$ 1$\mu$G, then assuming synchrotron radiation, the energy of particles emitting at $\sim$1\,keV would have to be $E[1{\rm keV}] \geq5\times10^{13}$\,eV, while  the particles energy associated with the voltage drop ($\Phi$) of the pulsar, is only 5$\times 10^{13}$\,eV, and a PWN can accelerate only a negligible fraction of particles beyond this energy. The observation of a bright X--ray nebula suggests that this cannot be the case. Despite this value of the magnetic field being smaller than typical values in the ISM we want to recall here that the PWN is not expanding in the ISM but inside the SNR ejecta (not to be confused with the SNR shell), whose magnetic filed can in principle be much smaller. One can also compute the value of the equipartition magnetic field. Given that most of the energy is injected in particles emitting in the soft--X--rays, equipartition must be computed with respect to them. The value for the equipartition magnetic field is found to be $B_{\rm pwn-eq}\sim 7-10\, \mu$G. Interestingly, such value is compatible with the two limits we found before. This suggests that, in the hypothesis of a synchrotron nebula, the magnetic field should then be in the range of a few$-20\mu$G.

Summarizing, this study shows that to explain the observed extended X--ray emission as a PWN:
\begin{enumerate}
\item  Most of the spin-down energy must be injected in particles.
\item  A large fraction ($25-30$\%) of the spin-down energy must be injected in particles emitting in the soft X--ray/EUV  ($E_b$=$E[0.3\,{\rm keV}]$).
\item The magnetic field in the nebula, $B_{\rm pwn}$, must be $<$ 20\,$\mu$G, otherwise there would be little energy left in the particles to explain the
observed X--ray efficiency.
\item $B_{\rm pwn}$ must be $>$ 1\,$\mu$G, otherwise one would need a large amount of particles with energies above the one given by the pulsar voltage drop $\Phi$ to explain the X--ray emission.
\item E[1\,keV] $\sim e\,\Phi$,  the energy related to pulsar voltage drop. This might explain the steep X--ray spectrum, with the X--ray 
emitting particles having energies close to the high-energy cutoff. 
\end{enumerate}

The above model was developed, assuming a standard PWN intepretation of the observed emission, where X--ray are due to synchrotron. 
However it is also possible that we are observing a Compton nebula, where the X--ray emission is due to Inverse Compton on the CMB, by a relic electron population with typical energies $\sim$1 GeV. For a magnetic field in the nebula $B_{\rm pwn}$ one would then expect a radio luminosity of $\sim10^{30} {\rm erg\; s}^{-1}(B_{\rm pwn}/3\mu{\rm G)^2}$. However a field $>10\mu$G is required for those particles to emit above the ionospheric cutoff ($\sim$30\,MHz). 

Another possible explanation of the extended emission is an exotic PWN, which might be powered by dissipation of magnetic field instead of the spin-down wind. One can try to estimate the value of the magnetic field at a typical distance from RRAT~J1819$-$1548, of the order of the size of the nebula. This depends on the dynamics with which the magnetic field is advected outward: continuous outflow vs sporadic bursts.

For a smooth continuous outflow  (either an unshocked flow, or for a strongly magnetized wind with a weak shock) the magnetic field at distances $\sim$R$_{\rm pwn}$, is found supposing a magnetic dipole from the stellar surface (R$_{psr}$) to the light cylinder (R$_{lc}$) and then a monopolar solution up to R$_{\rm pwn}$: \\
\begin{equation}
B(R_{\rm pwn}) \approx B_{\rm psr}(R_{\rm psr} / R_{\rm lc})^3(R_{\rm lc}/R_{\rm pwn}). \\
\end{equation}
which gives $B(R_{\rm pwn})\sim 0.5\,\mu$G.
This is a very small magnetic field, and as stated before particles at energy 
exceeding what is allowed by the pulsar voltage would be required to produce the observed X--ray emission. Moreover in a strongly magnetized outflow, only weak shocks are possible, and in this case shock acceleration is usually inefficient.

For a dynamical configuration, the nebular magnetic field might be provided by eruptive events related to fractures in the neutron star crust that take part of the magnetic energy stored in the neutron star and transfer it to the nebula. However for a bubble expanding adiabatically from a size of the order of the pulsar to a size of the order of the nebula the magnetic field should drop by a factor  $(R_{\rm psr}/R_{\rm pwn})^2 $ (magnetic flux conservation). Even for the very high inferred field at the neutron star surface, the values that are obtained at the distance of the nebula are far too small.

The above arguments appear to tentatively disfavour the magnetically powered idea, not just in the case of  RRAT~J1819$-$1548, but as a general interpretation of extended X--ray emission around strongly magnetized neutron stars. It is not a problem of explaining the luminosity (energetic), but of explaining radiation in the X--ray band as due to synchrotron. Interestingly  the magnetically powered idea could still work in the case of a Compton nebula, to energize the low energy relic electrons. 

The third possibility we want to discuss is that of a scattering halo. Scattering of X--rays by interstellar dust on the line of sight produces a scattered halo around the point source \cite[and references therein]{draine03}.  Therefore, for \f ~there might be a possibility that we are looking at a scattering halo. However prediction of scattering by diffuse dust in the ISM give halo sizes $\sim$ 10\,arcmin. Given the much smaller extension of the observed emission, reflection should be due to a local dust concentration around the source. We can estimate the flux of the halo using the method derived by \cite{draine04}  for direct determination of distances to nearby galaxies with bright  background AGNs, QSOs, or GRBs \citep[see e.g.][for other examples of X--ray scattering halos]{Rivera-Ingraham10}. The flux of the scattered photons, I$_{halo}$, is related to the flux in the point source, I$_{ptsrc}$, by I$_{ptsrc}$=(I$_{ptsrc}$+ I$_{halo}$)\,exp(-$\tau_{sca}$) \citep{draine03}, where the total scattering optical depth,  $\tau_{sca}$, may be determined using a model of interstellar dust consisting of a size distribution of carbonaceous and 
silicate grains: $\tau_{sca}$/A$_{\rm V}\approx$ 0.15\,(\textit{E}/keV)$^{-1.8}$ \citep{draine04}.  The optical extinction A$_{\rm V}$ 
may be obtained using the relation found by \cite{predehl95} (A$_{\rm V}$ = 0.56\,N$_{\rm H}$[10$^{21}$\,cm$^{-2}$]\,+\,0.23). For RRAT\,J1819--1458, using the parameters found in the present work (see Table 1) and the above equations, with $E\sim$1\,keV, we 
obtain A$_{\rm V}\sim$3.6 mag,  $\tau_{sca}\sim$0.54, I$_{ptsrc}\sim$1$\times$10$^{-13}$ erg\,cm$^{-2}$\,s$^{-1}$, 
and a I$_{halo}\sim$0.3$\times$10$^{-13}$\,erg\,cm$^{-2}$\,s$^{-1}$. This flux of the halo is of the order of  that observed for the extended X--ray emission (see Sect.\,3.3.2). 

This study has shown that the  extended X--ray emission around \f, detected with a very high significance,  if due to synchrotron requires a high efficiency of injection for the X--ray emitting particles, much higher than in young PWNe. If due to IC-CMB,  it could trace a relic population of pairs injected during the life of the system. Either a PWN  or a scattering halo are possible interpretations, while the magnetically powered scenario appears to be tentatively disfavoured in the case of synchrotron emission, while we cannot rule out that it can provide an energizing source in the case of a Compton nebula.

\vspace*{1.5cm}

{\textbf{Acknowledgments}. This work was supported by the grants AYA2009-07391 and SGR2009-811, as well as the Formosa program TW2010005 and iLINK program 2011-0303. NR is supported by a Ramon y Cajal research position in CSIC. PS
acknowledges support from NASA Contract NAS8-03060. This work was supported by Chandra Observer Support Award  GO1-12091X. This publication makes use of data products from the Two Micron All Sky Survey, which is a joint project of the University of Massachusetts and the Infrared Processing and Analysis Center/California Institute of Technology, funded by the National Aeronautics and Space Administration and the National Science Foundation.}


\end{document}